\PassOptionsToPackage{unicode}{hyperref}
\PassOptionsToPackage{hyphens}{url}
\PassOptionsToPackage{dvipsnames,svgnames,x11names}{xcolor}
\documentclass[
  conference]{IEEEtran}
\usepackage{xcolor}
\usepackage{amsmath,amssymb}
\usepackage{iftex}
\ifPDFTeX
  \usepackage[T1]{fontenc}
  \usepackage[utf8]{inputenc}
  \usepackage{textcomp} 
\else 
  \usepackage{unicode-math} 
  \defaultfontfeatures{Scale=MatchLowercase}
  \defaultfontfeatures[\rmfamily]{Ligatures=TeX,Scale=1}
\fi
\usepackage{lmodern}
\ifPDFTeX\else
\fi
\IfFileExists{upquote.sty}{\usepackage{upquote}}{}
\IfFileExists{microtype.sty}{
  \usepackage[]{microtype}
  \UseMicrotypeSet[protrusion]{basicmath} 
}{}
\usepackage{listings}
\newcommand{\passthrough}[1]{#1}
\usepackage{longtable,booktabs,array}
\usepackage{calc} 
\usepackage{etoolbox}
\makeatletter
\patchcmd\longtable{\par}{\if@noskipsec\mbox{}\fi\par}{}{}
\makeatother
\IfFileExists{footnotehyper.sty}{\usepackage{footnotehyper}}{\usepackage{footnote}}
\makesavenoteenv{longtable}
\usepackage{graphicx}
\makeatletter
\newsavebox\pandoc@box
\newcommand*\pandocbounded[1]{
  \sbox\pandoc@box{#1}%
  \Gscale@div\@tempa{\textheight}{\dimexpr\ht\pandoc@box+\dp\pandoc@box\relax}%
  \Gscale@div\@tempb{\linewidth}{\wd\pandoc@box}%
  \ifdim\@tempb\p@<\@tempa\p@\let\@tempa\@tempb\fi
  \ifdim\@tempa\p@<\p@\scalebox{\@tempa}{\usebox\pandoc@box}%
  \else\usebox{\pandoc@box}%
  \fi%
}
\def\fps@figure{htbp}
\makeatother
\usepackage{microtype}
\usepackage{booktabs}
\usepackage{caption}
\usepackage{listings}
\usepackage{xurl}
\usepackage{bookmark}
\IfFileExists{xurl.sty}{\usepackage{xurl}}{} 
\hypersetup{
  pdftitle={Above the Inner Loop: Exceeding Accelerate at LLM Prefill GEMM on the M1 AMX},
  pdfauthor={Deyvik Bhan},
  colorlinks=true,
  linkcolor={blue},
  filecolor={Maroon},
  citecolor={Blue},
  urlcolor={blue},
  pdfcreator={LaTeX via pandoc}}

\title{Above the Inner Loop: Exceeding Accelerate at LLM Prefill GEMM on
the M1 AMX}
\author{\IEEEauthorblockN{Deyvik Bhan}\IEEEauthorblockA{Georgia Institute of Technology\\dbhan6@gatech.edu}}
\date{}

\begin{document}
\maketitle
\begin{abstract}
On Apple Silicon CPUs the single-precision GEMMs that dominate large
language model (LLM) prefill are dispatched by Accelerate to a matrix
coprocessor -- the Apple Matrix Extension (AMX) on the M1 through M3,
the Arm Scalable Matrix Extension (SME) on the M4 and later. A recent
hand-written SME kernel, MpGEMM {[}1{]}, beats Accelerate by 1.23 on the
M4; we ask the analogous question on the older but far more widely
deployed M1 AMX and reach a structural conclusion: the speedup is not a
faster inner loop. By microbenchmark the M1 AMX inner loop is load-issue
bound -- once any operand load interleaves with the FMA32 stream,
single-thread throughput falls to a roughly 610-to-680 GFLOPS band,
under half the load-free rate, and none of the rearrangements we
evaluate escape it. The gain comes instead from two deployment-level
levers Accelerate underuses here: fine multi-thread panels that fill the
M1's second on-chip AMX block, winning the \(K \geq N\) shapes, and
pre-packing the constant weight at load, winning the \(N > K\) shapes. A
bit-exact direct-AMX kernel using both is the fastest bit-exact fp32
GEMM path we find on the M1: it exceeds the three general-purpose
Accelerate fp32 paths -- the callable cblas\_sgemm and BNNSMatMul and
the BNNS Graph compiler -- at all twelve LLM prefill GEMMs at
\(S = 128\) (GPT-2 to Llama-7B scale), leading the fastest Accelerate
path (BNNS Graph) by 1.17 -- and by 1.09 at the three shapes where BNNS
Graph also holds fp32 -- with a geometric mean of 1.58 over BNNSMatMul
and a larger margin (about \(2.0\times\)) over cblas\_sgemm. Every
output is bit-identical to Accelerate, whereas BNNS Graph is bit-exact
at only three of twelve shapes and computes the rest at reduced
precision (error up to 1.4e-3); even where both are exact the kernel
leads. Dropped into llama.cpp in place of its
\passthrough{\lstinline!cblas\_sgemm!} prefill matmuls, the kernel
raises measured full-forward throughput from 291 to 420 tokens per
second (\(1.44\times\), bit-identical) at the 128-token prefill regime
characterized here -- an end-to-end gain, not a GEMM-only ratio. The
contribution is this M1-AMX characterization -- load-issue
microbenchmark, two-block aggregate, a per-core occupancy probe showing
Accelerate idles the second AMX block at these shapes, and a
negative-results map -- leaving fine-panel scheduling and pre-packing as
the only two levers above an inner loop at the hardware limit;
mis-tuning the single column-panel width costs nearly \(2\times\).
\end{abstract}

\subsection{1. Introduction}\label{introduction}

On-device serving of LLMs on Apple Silicon CPUs is gated, during
prefill, by the throughput of the single-precision GEMMs (SGEMMs) that
constitute the dominant arithmetic work per transformer layer.
Accelerate is the standard SGEMM provider on these chips; internally it
dispatches SGEMM to a matrix coprocessor (AMX on the M1 through M3, SME
on the M4 and later). Because Accelerate exposes a single,
shape-agnostic entry point, its internal tile sizes are chosen to
perform reasonably across the full diversity of inputs rather than to
maximize throughput for any particular prefill GEMM shape.

Recent work has begun to quantify the resulting headroom, but mostly on
newer hardware. MpGEMM {[}1{]} shows that on the M4 Pro a hand-written
kernel using cache-aware partitioning, on-the-fly transposition, and
specialized SME microkernels exceeds Accelerate by a factor of 1.23
across DeepSeek and LLaMA prefill. On the M1 through M3 AMX path the
picture is less complete. Zhou's MEng thesis {[}2{]} reports an in-place
GEMM using masked outer products that exceeds Accelerate on the M2 at
the shapes evaluated there. The Apple vs.~Oranges study {[}3{]}
characterizes Accelerate's throughput on M1 through M4 without a
dedicated AMX kernel. The Bare-Metal Tensor Virtualization paper {[}4{]}
builds a from-scratch Arm64 LLM engine on NEON SIMD kernels rather than
the AMX matrix path, treating AMX as a ``black box''; on a
110M-parameter decode benchmark its hand-tuned NEON engine trails a
PyTorch + Accelerate (AMX) baseline by 4.87\(\times\) (61.3 against
298.7 tokens per second, computed from their reported throughputs). To
our knowledge no published work decomposes the M1 AMX gap to Accelerate
by LLM prefill GEMM shape class or across multiple model scales.

We ask, on the M1 AMX at the GEMM shapes of LLM prefill: where does the
throughput available to a hand-written kernel over Accelerate actually
come from? The answer is structural. The AMX inner loop is at its
hardware limit, so the gain is not in computing faster but in two
deployment-level choices Accelerate makes poorly at these shapes: it
leaves the second on-chip AMX block idle (measured directly, Section
4.6) and re-packs the constant weight on every call. The kernel
evaluated here applies three-level Goto--BLIS cache blocking with
explicit B-panel packing and four-way instruction-level parallelism
across the four AMX FMA32 banks, and pre-packs the constant weight once
at model load (Section 3.2); all twelve configurations are bit-identical
to Accelerate sgemm, and the kernel is released as a reusable component.

The work makes three contributions. \textbf{(1)} A micro-architectural
characterization of the M1 AMX inner loop: by isolated, cache-resident
microbenchmark (Section 4.2), throughput is load-issue bound -- with any
operand load interleaved with the FMA32 stream a single thread falls to
roughly the 610-to-680 GFLOPS band and stays there across the
output-tile shape, paired loads, the number of loads per iteration, the
load-to-FMA dependency, and phase batching, well under the load-free FMA
rate of about 1,525 GFLOPS. No single-thread rearrangement we evaluate
escapes the floor. \textbf{(2)} A bit-exact pre-packed direct-AMX kernel
that exceeds every Accelerate fp32 GEMM path we measure, with an
attribution of the gain. Fine multi-thread panels fill the second AMX
block Accelerate leaves idle and, on their own, win the \(K \geq N\)
shapes even when the kernel re-packs the weight on every call;
pre-packing the constant weight wins the \(N > K\) shapes. Neither lever
is itself new -- weight pre-packing is established practice in
production inference {[}14, 15{]}, and the M1's per-cluster AMX block
structure is documented in the reverse-engineered references {[}5, 6{]};
the contribution is the measurement that Accelerate forgoes both at the
\(M = 128\) prefill shapes, resolved by shape class and confirmed by a
direct per-core occupancy probe (Section 4.6). Together they give a
geometric mean of \(1.58\times\) over BNNSMatMul and \(1.17\times\) over
BNNS Graph at all twelve shapes (Sections 4.3, 4.5), every output
bit-identical to Accelerate. \textbf{(3)} A bounding of the M1-AMX
design space by negative results -- operand streaming, single-thread
microkernel restructuring, thread-count pinning, and 2-D output tiling
all leave throughput unmoved (Section 3.4) -- which leaves the
column-panel width as the one knob that matters, where mis-tuning a
single value costs nearly \(2\times\) (Section 4.6).

\subsection{2. Background}\label{background}

\subsubsection{2.1 LLM prefill GEMM
shapes}\label{llm-prefill-gemm-shapes}

A representative transformer block at prefill batch \(S = 128\), hidden
dimension \(H = 2{,}048\), and FFN multiplier 4 exhibits four GEMM shape
classes. Writing \((M, N, K)\) for the multiplication
\(C[M, N] = A[M, K] \cdot B[K, N]\), these are the QKV projection at
\((128, 2048, 2048)\), the FFN up-projection (FFN1) at
\((128, 8192, 2048)\), the FFN down-projection (FFN2) at
\((128, 2048, 8192)\), and -- once at the end of the model, not per
layer -- the LM head at \((128, V, 2048)\) with vocabulary \(V\)
(commonly 32,000 to 50,000 across recent open LLMs). The per-layer
attention-output projection is square \((128, H, H)\) and so falls in
the QKV shape class. These span the structural variation any prefill
GEMM kernel must accommodate: square (QKV and the attention-output
projection), N-large (FFN1), K-large (FFN2), and N-very-large (LM head).
Table 3 evaluates twelve such shapes across three model scales. The
``GPT-2-style'' row group is an \(H = 2{,}048\) configuration with FFN
multiplier 4 and a large vocabulary (\(V = 60{,}000\)), used to populate
the per-shape evaluation with a large-vocabulary LM head case; it is a
label for that group, not a reproduction of GPT-2-small (\(H = 768\)).

\subsubsection{2.2 Apple AMX}\label{apple-amx}

The Apple Matrix Extension is a coprocessor that shares the L2 cache of
its P-cluster. Operations are encoded as \passthrough{\lstinline!.word!}
directives following a reserved Arm64 instruction; the encodings are not
part of Apple's public ISA, and we use the reverse-engineered references
{[}5, 6{]}. On M1 the fp32 GEMM primitives are AMX\_LDX (load 64 or 128
bytes into one or two X registers, pair bit at 62), AMX\_LDY (load 64
bytes into a Y register), AMX\_FMA32 (a 16×16 outer-product
fused-multiply-add \(Z[j][i] \mathrel{+}= X[i] \cdot Y[j]\), skip-Z
control at bit 27, four Z banks indexed at bits 20--22), and AMX\_STZ
(store one Z row, 64 bytes). The M1 contains one AMX block per cluster:
one shared by the four P-cores and a second, lower-throughput block
shared by the four E-cores (Section 4.6). A cache-resident
microbenchmark issuing only outer products reaches about 1,525 GFLOPS
fp32 on one thread, but with operand loads interleaved a single thread
drops into roughly the 610-to-680 GFLOPS band (Section 4.2); multiple
threads sharing the two blocks recover an aggregate near 1,480 GFLOPS
(Section 4.6). Accelerate sgemm dispatches to AMX internally.

\subsection{3. Method}\label{method}

\subsubsection{3.1 Three-level cache
blocking}\label{three-level-cache-blocking}

The microkernel adopts the Goto--BLIS three-level blocking structure
{[}7{]}, adapted to the M1 cache hierarchy (a 128 KB L1 data cache per
core, a 12 MB L2 shared across the P-cluster, no L3). Matrix A is
pre-transposed once into row-major At{[}K, M{]} so that a 16-element
column slice of A loads contiguously into a single AMX Y register;
matrix B is packed into row-major contiguous buffers packB{[}K\_c,
N\_c{]}. With \(M_r = 16\) the AMX outer-product height and \(N_r = 64\)
the per-iteration output panel width (four Z banks of 16 lanes each),
the blocked GEMM is:

\begin{figure*}[t]
\centering
\begin{minipage}{0.78\textwidth}
\begin{lstlisting}
for jc in 0..N step Nc:                                # B panel L2-resident
  for pc in 0..K step Kc:                              # A panel L1-resident
    pack B[pc:pc+Kc, jc:jc+Nc] into packB[Kc, Nc]
    for i0 in 0..M step Mr=16:
      for jr in 0..Nc step Nr=64:
        if pc > 0:                                     # carry partial Z from C
          for t in 0..3, j in 0..15:
            AMX_LDZ(C[i0+j, jc+jr+16*t] → Z row 4*j+t)
        for kk in 0..Kc:
          AMX_LDY(At[pc+kk, i0:i0+16] → Y[0])
          AMX_LDX(packB[kk, jr:jr+32]  → X[0,1])       # LDX_pair
          AMX_LDX(packB[kk, jr+32:jr+64] → X[2,3])     # LDX_pair
          first = (pc==0 and kk==0)
          AMX_FMA32(bank 0, x_off=  0, first)          # 4-way ILP
          AMX_FMA32(bank 1, x_off= 64, first)
          AMX_FMA32(bank 2, x_off=128, first)
          AMX_FMA32(bank 3, x_off=192, first)
        for t in 0..3, j in 0..15:
          AMX_STZ(Z row 4*j+t → C[i0+j, jc+jr+16*t])   # always STZ
\end{lstlisting}
\end{minipage}
\end{figure*}

The microkernel reads four X registers of B (two LDX\_pair) and one Y
register of A, and issues four FMA32 across the four Z banks. The
Z-carry is correctness-critical: for \(pc > 0\) the partial accumulator
is reloaded from C with LDZ at the start of each \((i0, jr)\) tile and
stored back with STZ at the end of every iteration, and the skip-Z flag
is asserted only at \(pc = 0\), \(kk = 0\). Without this discipline the
Z state of one tile leaks into the next across \(pc\) iterations and
produces silent numerical drift.

\subsubsection{3.2 Weight pre-packing}\label{weight-pre-packing}

In LLM inference the second operand B is a model weight: constant across
every token, prefill, and decode step of a session. A stateless GEMM
interface re-packs it on every call; a specialized kernel can instead
pack each weight once at model load and reuse the packed form for the
process lifetime. Pre-packing constant weights is standard in production
inference libraries {[}14, 15{]}, but the classic Accelerate entry
points do not amortize it: cblas\_sgemm and BNNSMatMul receive the
weight per call, and the BNNS fully-connected filter, which does accept
a weight at construction, packs it into a path Section 4.3 shows is
slower than this kernel at the large-\(N\) shapes where we measure it.
(Apple's BNNS Graph also repacks automatically {[}16{]}; Section 4.5
compares directly.) The kernel packs each weight once into the
\([K, N_c]\) panels of Section 3.1 at load time and, on each call,
executes only the LDY/LDX/FMA32/LDZ/STZ compute loop against the
resident panels. An engine that stores weights as \(W[N, K]\) -- as
llama.cpp does, passing \passthrough{\lstinline!CblasTrans!} -- must
transpose once at load, a one-time cost under pre-packing (Section 4.7).
The pack is paid once and amortized to zero, and the per-call arithmetic
is bit-identical to Accelerate. This amortization is the source of the
win at the \(N > K\) shapes; at the \(K \geq N\) shapes the kernel wins
even without it, on the multi-thread panel granularity of Section 4.6
(Section 4.4).

\subsubsection{3.3 Panel sizing}\label{panel-sizing}

With the weight pre-packed, the approximately 480 MB pack write at the
LM head shape is paid once at load, so the pre-packed kernel runs the
packed compute loop at every shape. Two panel parameters remain: the
column-panel width \(N_c\), which sets the multi-threading granularity,
and the K-blocking depth \(K_c\), which sets the Z-accumulator reload
frequency (a larger \(K_c\) runs more inner iterations before the
partial sum is stored to and reloaded from C). An offline sweep over
\(N_c \in \{64, 128, 192, 256, 384, 512\}\) and
\(K_c \in \{256, 512, 1024, 2048\}\), rejecting any selection not
bit-identical to Accelerate, settles the deployed pair at \(N_c = 64\),
\(K_c = 2{,}048\). No single pair is each shape's individual optimum
(the \(K = 2{,}048\) square shapes prefer \(K_c = 1{,}024\) by about two
percent), but \(N_c = 64\), \(K_c = 2{,}048\) wins all twelve shapes
against every Accelerate path, bit-exact, and the kernel uses it
uniformly; it also raises the geometric mean over BNNSMatMul (1.56 at
\(K_c = 1{,}024\) to 1.58), so it is not tuned against any one
comparison. \(N_c = 64\) produces one panel per core and so reaches both
AMX blocks (Section 4.6); \(K_c = 2{,}048\) is affordable only because
the weight is pre-packed -- it halves the reload passes at large \(N\),
but a per-call kernel cannot afford to pack a 2,048-deep panel on every
call (Section 4.4). At \(K_c = 2{,}048\) the per-tile A-slice
\(At[pc{:}pc{+}K_c,\, i_0{:}i_0{+}16]\) is exactly
\(2{,}048 \times 16 \times 4 = 128\) KB, sizing it to the M1 P-core L1
data cache, which is why deeper blocks gain nothing. The \(N_c\) choice
is a tuning constant fixed by the offline sweep: a coarse \(N_c = 512\)
reaches only one AMX block, so the deployed kernel uses the finer
\(N_c = 64\) (Section 4.6).

\subsubsection{3.4 Negative results}\label{negative-results}

Four interventions -- drawn from the AMX instruction surface, general
latency-hiding practice, and the M1 multi-core layout -- were evaluated;
none improves on the pre-packed kernel at the right panel width (Table
1). Together they bound the headroom: the inner loop is at the issue
limit, and the only levers that move throughput are removing per-call
work (pre-packing) and a panel width fine enough to keep both AMX blocks
busy. Two deserve a specific note. The LDX\_pair encoding -- a 128-byte
paired load halving the per-\(k\) load count -- does not raise
throughput, consistent with the load-issue account of Section 4.2 in
which load count is shown not to bound the inner loop. Software PRFM
prefetch ahead of LDX changes nothing within run-to-run noise: the M1
hardware prefetcher already detects the sequential B-row stride.

\begin{table}[t]
\centering
\caption*{Table 1. Interventions evaluated and rejected; none exceeds the
pre-packed kernel of Table 3.}
\footnotesize
\setlength{\tabcolsep}{4pt}
\begin{tabular}{@{}
>{\raggedright\arraybackslash}p{(\linewidth - 4\tabcolsep) * \real{0.3333}}
  >{\raggedright\arraybackslash}p{(\linewidth - 4\tabcolsep) * \real{0.3333}}
  >{\raggedright\arraybackslash}p{(\linewidth - 4\tabcolsep) * \real{0.3333}}@{}}

\toprule
\begin{minipage}[t]{\linewidth}\raggedright
intervention
\end{minipage} & \begin{minipage}[t]{\linewidth}\raggedright
outcome
\end{minipage} & \begin{minipage}[t]{\linewidth}\raggedright
section
\end{minipage} \\
\midrule

operand streaming (unpacked, no B pack) & 0.15--0.5× of packed;
cache-hostile B access & §4.4 \\
microkernel restructuring (tile, pairing, pipelining, batching) & no
change; inner loop is issue-bound & §4.2 \\
explicit thread-count pinning & within noise of default GCD at
\(N_c = 64\) & §4.6 \\
2-D output tiling (split the \(M\) dimension) & lowers throughput; AMX
wants the full \(M = 128\) per work item & §4.6 \\
\bottomrule
\end{tabular}
\end{table}

\subsection{4. Evaluation}\label{evaluation}

\subsubsection{4.1 Experimental setup}\label{experimental-setup}

Measurements are on an Apple M1 (four P-cores at a 3.228 GHz nominal
maximum, four E-cores) under macOS Tahoe 26.5.1, compiled with Apple
clang 21.0.0 at \passthrough{\lstinline!-O3!}. Each configuration is the
median over eleven isolated binary invocations; within an invocation a
kernel is timed over at least fifteen trials after warm-up and the
per-invocation median taken. The proposed kernel dispatches through
Grand Central Dispatch and holds no persistent worker pool, so the
backends can be timed in turn within one process; the BNNS Graph
comparison (Section 4.5) runs in a separate binary. Processes run
unpinned under the default scheduler. All results are from a single M1
unit (T8103); we do not control for silicon-to-silicon variation, and
the absolute GFLOPS figures carry this part's thermal range (the
load-free inner loop spans 900 to 1,600 GFLOPS, Section 4.2). The
speedup ratios are formed within a single process per invocation, so a
thermal excursion scales every backend together and largely cancels in
the ratio.

Numerical correctness is verified against Accelerate sgemm by sampling
the full output C at a large coprime stride (every 997th or
\(1{,}023\)rd element, so the sample sweeps all rows and columns); every
bit-exact configuration satisfies
\passthrough{\lstinline!max-abs-diff = 0e+00!} over that sample. The
headline measurements run each backend at its default threading in a
dedicated process -- the comparison a deployed serving stack observes --
rather than constraining Accelerate to a single thread. The backends are
cblas\_sgemm, BNNSMatMul, the BNNS fully-connected filter (BNNS-FC,
reported as context only, Section 4.3), BNNS Graph (Section 4.5), and
the proposed pre-packed kernel. As an external reference, OpenBLAS sgemm
0.3.33 (neoversen1 NEON) sustains roughly 200 to 300 GFLOPS on the M1
P-core, confirming all AMX backends operate in a regime NEON cannot
reach (we verified this OpenBLAS build does not dispatch to Accelerate).
BNNS Graph requires macOS 15 or later; it is measured on the same Tahoe
26.5.1 release as every other backend here.

\subsubsection{4.2 The throughput ceiling and the load-issue
bound}\label{the-throughput-ceiling-and-the-load-issue-bound}

Two ceilings frame every result that follows. A cache-resident
microbenchmark issuing only FMA32 outer products, with no loads,
sustains about 1,525 GFLOPS on one thread (900 to 1,600 across thermal
states) -- unattainable by any GEMM, which must load its operands. With
even one operand load interleaved into the FMA32 stream, single-thread
throughput collapses to roughly the 610-to-680 GFLOPS band and stays
there (Table 2): it moves little with the number of loads per iteration,
paired versus single loads, whether the loaded register feeds the
following FMA, the output-tile shape (\(16\times64\) versus
\(32\times32\)), or batching four iterations' loads ahead of issue -- in
every case under half the load-free rate. The invariance is consistent
with loads and FMA32 contending for a single AMX issue slot, while the
\(1{,}024\)-element Z accumulator caps the work at four FMA32 per load
group; no rearrangement we tried raises the compute issued per load. All
rows of Table 2 are issued back-to-back within a single process at a
matched thermal state, so the no-load-to-loaded collapse is a
within-session contrast and the broad thermal range of Section 4.1 does
not enter the gap. This is the basis for the per-call result of Section
4.4.

\begin{table}[t]
\centering
\caption*{Table 2. Single-thread throughput of the AMX inner loop as the
load-to-FMA32 interleaving varies, all operands L1-resident so the
measurement is issue-bound. Each row issues four FMA32 per iteration
(2,048 FLOP); the loads column counts AMX load instructions interleaved
with them. Source:
\protect\texttt{bench/amx/amx\_microbench.cc}.}
\footnotesize
\setlength{\tabcolsep}{4pt}
\begin{tabular}{@{}
lll@{}}

\toprule
inner loop & loads : FMA32 & GFLOPS \\
\midrule

FMA32 only (no loads) & 0 : 4 & \textasciitilde1,525 \\
+ 1 paired load & 1 : 4 & 677 \\
+ 1 paired load, no data dependency & 1 : 4 & 679 \\
+ 2 paired loads (\(32\times32\) tile) & 2 : 4 & 609 \\
+ 3 loads (\(16\times64\) tile, deployed) & 3 : 4 & 643 \\
4-iteration phase batch & 8 : 16 & \textasciitilde610 \\
\bottomrule
\end{tabular}
\end{table}

The single-thread floor is not the multi-thread ceiling. At the large
shapes of Table 3 several threads share one block, one thread's loads
issuing while another's FMA32 occupy it, and the cache-resident
aggregate rises to roughly 1,480 GFLOPS at eight threads (Table 5). With
the weight pre-packed at a fine enough panel granularity (Section 4.6)
the proposed kernel sustains 1,125 to 1,250 GFLOPS on the real shapes,
well above the single-thread floor.

\subsubsection{4.3 Main result: twelve LLM prefill
GEMMs}\label{main-result-twelve-llm-prefill-gemms}

Table 3 reports per-shape throughput for the twelve prefill GEMMs across
three model scales (a GPT-2-style group at \(H = 2{,}048\) with a
large-vocabulary LM head, TinyLlama-1.1B at \(H = 2{,}048\), and
Llama-7B at \(H = 4{,}096\)) at \(S = 128\), against the two callable
Accelerate fp32 entry points -- cblas\_sgemm and BNNSMatMul -- both of
which re-pack the weight on every call. BNNS Graph, the ahead-of-time
graph compiler, is a separate category compared in Section 4.5.

\begin{figure}[t]
\centering
\centering
\pandocbounded{\includegraphics[keepaspectratio,alt={Prefill GEMM throughput (fp32, bit-exact) at the GPT-2-style scale (H = 2\{,\}048): the proposed pre-packed kernel against the two callable Accelerate fp32 paths, cblas\_sgemm and BNNSMatMul, both of which re-pack the weight on every call. Bars are median GFLOPS over eleven isolated invocations (the GPT-2-style rows of Table 3). The kernel leads at all four prefill shape classes -- square (QKV), N-large (FFN-up), K-large (FFN-down), and N-very-large (LM head).}]{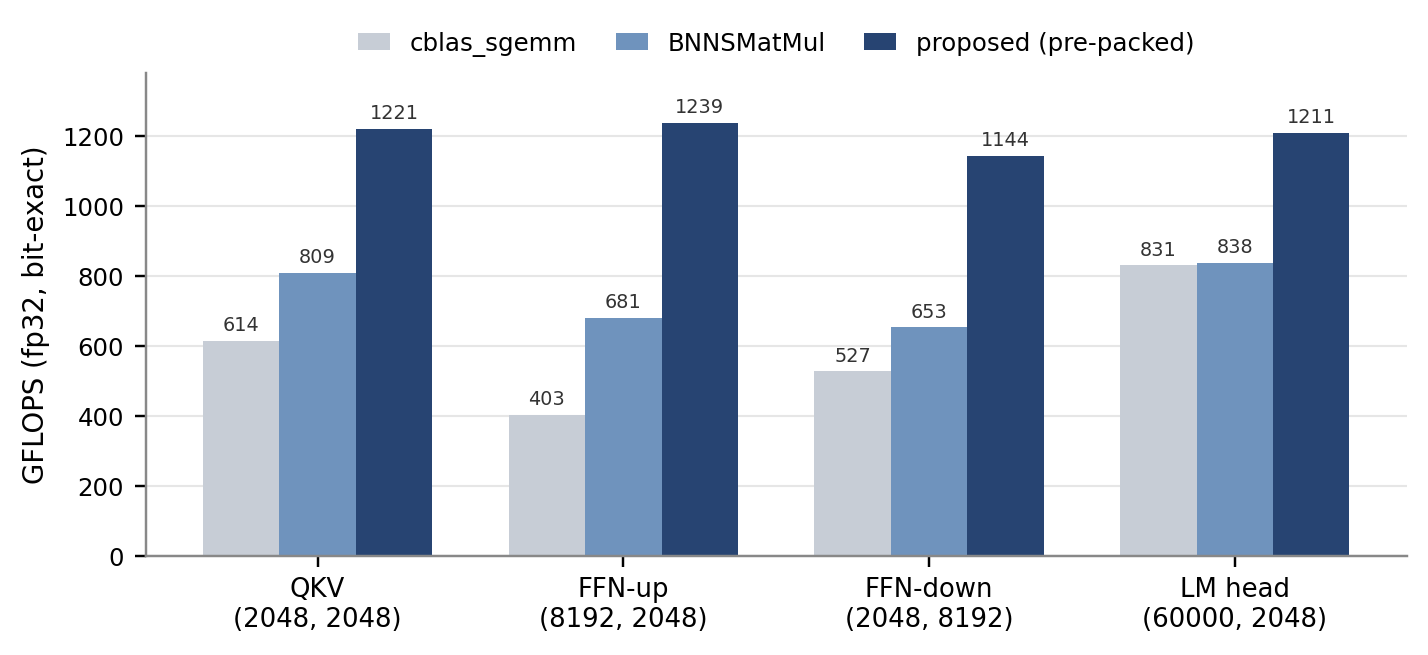}}
\caption{Prefill GEMM throughput (fp32, bit-exact) at the GPT-2-style
scale (\(H = 2{,}048\)): the proposed pre-packed kernel against the two
callable Accelerate fp32 paths, cblas\_sgemm and BNNSMatMul, both of
which re-pack the weight on every call. Bars are median GFLOPS over
eleven isolated invocations (the GPT-2-style rows of Table 3). The
kernel leads at all four prefill shape classes -- square (QKV),
\(N\)-large (FFN-up), \(K\)-large (FFN-down), and \(N\)-very-large (LM
head).}
\end{figure}

\begin{table*}[t]
\centering
\caption*{Table 3. Throughput at \(S = 128\) on the M1, in GFLOPS (every
row has \(M = S = 128\); shapes listed as \((N, K)\)), median over
eleven isolated binary invocations, all backends timed in one process
per invocation. cblas\_sgemm and BNNSMatMul re-pack the weight on every
call; the proposed kernel pre-packs it once at model load. The final
column is the ratio of the proposed kernel to BNNSMatMul. Every
proposed-kernel cell is bit-identical to Accelerate
(\protect\texttt{max-abs-diff = 0e+00}).
``GPT-2-style'' denotes an \(H = 2{,}048\) configuration with FFN
multiplier 4 and vocabulary \(60{,}000\), not the actual GPT-2-small
architecture (\(H = 768\)).}
\footnotesize
\setlength{\tabcolsep}{4pt}
\begin{tabular}{@{}
lllllll@{}}

\toprule
model & op & \((N, K)\) & cblas & BNNSMatMul & proposed & / BNNS \\
\midrule

GPT-2-style & QKV & (2048, 2048) & 614 & 809 & 1221 & 1.51 \\
GPT-2-style & FFN1 (up) & (8192, 2048) & 403 & 681 & 1239 & 1.82 \\
GPT-2-style & FFN2 (down) & (2048, 8192) & 527 & 653 & 1144 & 1.75 \\
GPT-2-style & LM head & (60000, 2048) & 831 & 838 & 1211 & 1.45 \\
TinyLlama-1.1B & QKV & (2048, 2048) & 614 & 800 & 1214 & 1.52 \\
TinyLlama-1.1B & FFN1 & (5632, 2048) & 839 & 795 & 1230 & 1.55 \\
TinyLlama-1.1B & FFN2 & (2048, 5632) & 520 & 680 & 1173 & 1.73 \\
TinyLlama-1.1B & LM head & (32000, 2048) & 899 & 850 & 1252 & 1.47 \\
Llama-7B & QKV & (4096, 4096) & 369 & 725 & 1236 & 1.70 \\
Llama-7B & FFN1 & (11008, 4096) & 804 & 830 & 1246 & 1.50 \\
Llama-7B & FFN2 & (4096, 11008) & 336 & 697 & 1125 & 1.61 \\
Llama-7B & LM head & (32000, 4096) & 735 & 843 & 1216 & 1.44 \\
\bottomrule
\end{tabular}
\end{table*}

The pre-packed kernel exceeds BNNSMatMul at all twelve (geometric mean
\(1.58\times\), range \(1.44\) to \(1.82\)) and cblas\_sgemm at all
twelve by a larger margin (geometric mean \(2.04\times\)). The advantage
is neither a faster inner loop nor pre-packing alone: Section 4.4
separates the two -- at the \(K \geq N\) shapes the kernel beats
BNNSMatMul even when re-packing the weight on every call, on the panel
granularity of Section 4.6; at the \(N > K\) shapes the win is
amortizing the weight pack. Accelerate's other callable pre-packing
path, BNNS-FC, captures neither: it collapses to 467 to 478 GFLOPS at
the large-\(N\) Llama shapes where the proposed kernel sustains above
\(1{,}100\). We measured BNNS-FC only there, where its construction-time
pre-pack is most favorable, and report it as context. The ``exceeds
every Accelerate fp32 GEMM path'' claim thus holds at all twelve shapes
for cblas\_sgemm, BNNSMatMul, and BNNS Graph (Section 4.5), and at the
large-\(N\) shapes for BNNS-FC.

The margin is not uniform across the table. The proposed kernel sustains
a nearly shape-independent 1,125 to 1,252 GFLOPS, so its margin over
BNNSMatMul is essentially the inverse of BNNSMatMul's own throughput.
That throughput is lowest at the FFN down-projection (\(K > N\), the
\(K\)-heavy small-\(N\) shape: 653 to 697 GFLOPS), giving the largest
margin (geometric mean \(1.70\times\)), and highest at the large-\(N\)
LM head (\(N \gg K\), 838 to 850 GFLOPS), giving the smallest
(\(1.45\times\)); the ordering tracks the baseline's absolute
throughput, not the relative magnitudes of \(K\) and \(N\). Across the
eleven invocations the proposed kernel is also the most reproducible of
the three backends in the worst case (throughput coefficient of
variation \(\leq 3.6\) percent at every shape, against 12 percent for
cblas at the small QKV shape and 5.1 percent for BNNSMatMul), so its
tail-latency behaviour is at least as predictable as Accelerate's.

\subsubsection{4.4 Per-call comparison and
ablation}\label{per-call-comparison-and-ablation}

Two questions remain: how the kernel compares when both re-pack the
weight on every call (isolating the inner loop from the pre-packing
advantage), and how that inner loop was built.

\textbf{Per-call.} The multi-threaded kernel without pre-packing,
re-tuned to its own best depth (\(K_c = 1{,}024\); the deeper
\(K_c = 2{,}048\) is too costly to pack on every call), exceeds
BNNSMatMul at seven of the twelve shapes (geometric mean 1.07). All six
\(K \geq N\) shapes -- the QKV projections and the FFN down-projections
-- win by 1.05 to 1.46, where the per-call pack is cheap relative to
compute and the fine panel granularity of Section 4.6 decides it. Of the
six \(N > K\) shapes, five fall behind at 0.87 to 0.92 because packing
the large-\(N\) weight on every call dominates; the GPT-2-style FFN
up-projection sits at parity (1.03). So multi-thread panel granularity
alone -- the first lever -- already wins the \(K \geq N\) shapes.

\textbf{Pre-packing}, the second lever, removes the per-call pack
(converting the \(N > K\) shapes to wins) and, because the pack is now
paid once at load, unlocks a deeper \(K_c = 2{,}048\) that halves the
Z-accumulator reload passes (Section 3.3) -- a depth the per-call kernel
cannot afford (at \(K_c = 2{,}048\) it wins only four of twelve). The
two levers are therefore not independent: the deeper block depth is a
benefit of pre-packing, so the per-call comparison is taken at the
configuration each mode runs best. Together they lift the geometric mean
to 1.58 over BNNSMatMul and win all twelve against every Accelerate
path, including BNNS Graph (Tables 3, 4).

\textbf{Inner loop.} Two single-thread choices account for what
inner-loop gain there is, both standard BLIS techniques on a
non-standard instruction set: explicit B-panel packing (retaining
\(B[:, jc{:}jc{+}N_c]\) in L2 across the \(i\) sweep) and \(K_c\)
blocking with LDZ/STZ carry of Z across \(pc\) (retaining
\(A[:, pc{:}pc{+}K_c]\) in L1). The work was selecting \((N_c, K_c)\)
and establishing the bit-exact carry, not inventing a technique; the
interventions closer to the instruction surface (LDX\_pair, software
pipelining) neither help nor hurt, consistent with Section 4.2. Skipping
the pack entirely and streaming B unpacked is the counterfactual: it
falls to 0.15 to 0.5 times the packed kernel as the strided B access
turns cache-hostile (Table 1), which is why the pack is retained.

\textbf{Batch size.} As a sensitivity check, the per-call kernel
(re-packing each call, single thread) was measured at the QKV shape
across \(S \in \{16, 32, 64, 128, 256, 512\}\) against single-thread
cblas\_sgemm: it does not exceed cblas at any batch size (ratio 0.54 to
0.70), worst at \(S = 512\) where its own throughput drops as the
transpose and pack working set spills the cache while cblas keeps
scaling. This confirms the headline advantage is a property of the
pre-packed deployment, not of the inner kernel at any batch size.

\subsubsection{4.5 Comparison with BNNS
Graph}\label{comparison-with-bnns-graph}

The comparison a reviewer asks for is against BNNS Graph, the macOS 15
path that performs automatic weight repacking. It is not a callable GEMM
but an ahead-of-time graph compiler that consumes a serialized model
(\passthrough{\lstinline!.mlmodelc!}) and repacks any constant weight at
compile time. Per shape we emit a single-operation graph \(y = xW\) with
\(W\) the constant weight, force fp32 storage and compute, and compile
once -- the compile, which performs the repack, is untimed, matching the
amortized pre-pack of the proposed kernel. Only graph execution is
timed; because BNNS Graph runs in its own binary (Section 4.1) rather
than alongside the other backends, we take the per-invocation median
over thirty graph calls instead of the fifteen used in the
shared-process harness, then the median over eleven invocations as
elsewhere. The output is checked against cblas\_sgemm.

\begin{table*}[t]
\centering
\caption*{Table 4. The proposed kernel versus BNNS Graph at \(S = 128\)
on the M1, GFLOPS (median over eleven invocations). ``BG max-diff'' is
the maximum absolute difference of the BNNS Graph output from
cblas\_sgemm; the proposed kernel's own difference is 0 at every shape.
BNNS Graph is bit-exact only at the three square QKV shapes; at the nine
rectangular shapes it dispatches to a reduced-precision
kernel.}
\footnotesize
\setlength{\tabcolsep}{4pt}
\begin{tabular}{@{}
>{\raggedright\arraybackslash}p{(\linewidth - 12\tabcolsep) * \real{0.1429}}
  >{\raggedright\arraybackslash}p{(\linewidth - 12\tabcolsep) * \real{0.1429}}
  >{\raggedright\arraybackslash}p{(\linewidth - 12\tabcolsep) * \real{0.1429}}
  >{\raggedright\arraybackslash}p{(\linewidth - 12\tabcolsep) * \real{0.1429}}
  >{\raggedright\arraybackslash}p{(\linewidth - 12\tabcolsep) * \real{0.1429}}
  >{\raggedright\arraybackslash}p{(\linewidth - 12\tabcolsep) * \real{0.1429}}
  >{\raggedright\arraybackslash}p{(\linewidth - 12\tabcolsep) * \real{0.1429}}@{}}

\toprule
\begin{minipage}[t]{\linewidth}\raggedright
model
\end{minipage} & \begin{minipage}[t]{\linewidth}\raggedright
op
\end{minipage} & \begin{minipage}[t]{\linewidth}\raggedright
proposed
\end{minipage} & \begin{minipage}[t]{\linewidth}\raggedright
BNNS Graph
\end{minipage} & \begin{minipage}[t]{\linewidth}\raggedright
/ BG
\end{minipage} & \begin{minipage}[t]{\linewidth}\raggedright
BG max-diff
\end{minipage} & \begin{minipage}[t]{\linewidth}\raggedright
BG bit-exact
\end{minipage} \\
\midrule

GPT-2-style & QKV & 1221 & 1116 & 1.09 & 0 & yes \\
GPT-2-style & FFN1 & 1239 & 983 & 1.26 & 2.5e-4 & no \\
GPT-2-style & FFN2 & 1144 & 812 & 1.41 & 9.3e-4 & no \\
GPT-2-style & LM head & 1211 & 1021 & 1.19 & 2.8e-4 & no \\
TinyLlama-1.1B & QKV & 1214 & 1116 & 1.09 & 0 & yes \\
TinyLlama-1.1B & FFN1 & 1230 & 1014 & 1.21 & 2.4e-4 & no \\
TinyLlama-1.1B & FFN2 & 1173 & 1057 & 1.11 & 5.8e-4 & no \\
TinyLlama-1.1B & LM head & 1252 & 1037 & 1.21 & 2.7e-4 & no \\
Llama-7B & QKV & 1236 & 1135 & 1.09 & 0 & yes \\
Llama-7B & FFN1 & 1246 & 1132 & 1.10 & 4.9e-4 & no \\
Llama-7B & FFN2 & 1125 & 911 & 1.23 & 1.4e-3 & no \\
Llama-7B & LM head & 1216 & 1115 & 1.09 & 5.5e-4 & no \\
\bottomrule
\end{tabular}
\end{table*}

The kernel is faster than BNNS Graph at all twelve shapes (geometric
mean \(1.17\times\), range \(1.09\) to \(1.41\)) and bit-exact where
BNNS Graph often is not. BNNS Graph matches cblas\_sgemm exactly only at
the three square QKV shapes; at every rectangular shape it differs by
\(2.4 \times 10^{-4}\) to \(1.4 \times 10^{-3}\), roughly a hundred
times the fp32 rounding floor, identical to the last digit across all
eleven invocations -- so it computes the rectangular GEMMs at reduced
precision despite fp32 storage. That BNNS Graph spends precision for
speed at the rectangular shapes points to where the remaining AMX
throughput is; a mixed-precision kernel could make the same trade, which
we leave to follow-up work. The closest cases are the QKV shapes, a
clean bit-exact-against-bit-exact \(1.09\times\), and the Llama LM head
(\(1.09\times\)), the rectangular shape nearest the inner loop -- even
there the proposed kernel is faster and bit-exact while BNNS Graph is
not. The \(1.09\times\) at the three QKV shapes is small but not within
run-to-run noise: across the eleven isolated invocations the proposed
kernel's slowest run exceeds BNNS Graph's fastest at every one of the
three (the two backends' per-invocation distributions do not overlap),
so the margin is resolved despite its size.

\subsubsection{4.6 Multi-thread scaling and the shared AMX
block}\label{multi-thread-scaling-and-the-shared-amx-block}

The single-thread floor (610-to-680 GFLOPS, Section 4.2) is not the
multi-thread ceiling. The M1 P-cluster has one AMX block shared by its
four cores, and one thread's loads issue while another's FMA32 occupy
the block. A cache-resident microbenchmark in which \(T\) threads drive
the shared block shows loaded throughput rising from 673 at one thread
to the P-cluster's pure-FMA ceiling -- it peaks at three threads (1,373)
and holds the roughly 1,330 rate at four (1,324 against a 1,332 pure-FMA
aggregate, Table 5), the 3-to-4-thread step within run-to-run spread --
so by four threads the loads are fully hidden behind cross-thread
compute. Extending past four onto the E-cluster's second block lifts the
aggregate to 1,483 at eight threads.

\begin{table}[t]
\centering
\caption*{Table 5. Aggregate throughput of the on-chip AMX blocks as the
thread count varies, cache-resident (issue-bound). loaded-MT runs the
full load + FMA32 inner loop; pure-FMA-MT issues no loads. Threads
beyond four spill from the P-cluster onto the E-cluster block. Source:
\protect\texttt{bench/amx/amx\_mt\_ceiling.cc}.}
\footnotesize
\setlength{\tabcolsep}{4pt}
\begin{tabular}{@{}
lll@{}}

\toprule
threads & loaded-MT (GFLOPS) & pure-FMA-MT (GFLOPS) \\
\midrule

1 & 673 & 1,528 \\
2 & 1,219 & 1,419 \\
3 & 1,373 & 1,418 \\
4 & 1,324 & 1,332 \\
6 & 1,435 & 1,542 \\
8 & 1,483 & 1,581 \\
\bottomrule
\end{tabular}
\end{table}

The two columns isolate why multi-threading is the lever and the inner
loop is not. The pure-FMA aggregate barely rises with thread count --
1,528 at one thread (matching the roughly 1,525 microbenchmark of
Section 4.2) to 1,581 at eight, +3 percent net over a non-monotonic
1,332-to-1,581 range -- because one thread's FMA32 stream already
saturates the issue port, so adding threads buys nothing without loads
to hide. The loaded aggregate rises 2.2-fold over the same range (673 to
1,483), entirely because each thread's loads slot into the issue gaps
the others' FMA32 leave. The gap between the columns is the load-issue
bound of Section 4.2 made visible across threads.

The lever that reaches this aggregate on the real shapes is the panel
granularity. The kernel parallelizes over column panels of width
\(N_c\): there are \(\lceil N/N_c \rceil\) of them, and Grand Central
Dispatch can place only as many threads as panels. At the QKV shapes
(\(N = 2{,}048\)) a coarse \(N_c = 512\) yields four panels -- enough to
occupy the P-cluster but not to reach the E-cluster block, and wide
enough that each panel's shared-L2 footprint thrashes the cache -- so
the kernel stalls near 630 GFLOPS, well below both the four-thread
cache-resident aggregate of Table 5 (1,324) and both Accelerate
baselines (Figure 2). Shrinking to \(N_c = 64\) produces 32 panels,
fills all eight cores and both blocks, and lifts the same shape to
roughly 1,200 GFLOPS, past BNNSMatMul and BNNS Graph alike, bit-exact.
Measured at the small-to-mid-\(N\) QKV and FFN-down shapes of all three
scales (median over eleven invocations, bit-exact,
\nolinkurl{bench/amx/amx\_retune\_gain.cc}), shrinking
from \(N_c = 512\) to \(N_c = 64\) lifts throughput by 85 to 102 percent
-- a geometric mean near \(1.9\times\). The gain persists even at the
\(N = 4{,}096\) shapes, whose eight \(N_c = 512\) panels already fill
all eight cores, so finer panels help through more than core coverage: a
512-wide packed panel also carries an eightfold larger shared-L2
footprint than a 64-wide one, so panel count and L2 footprint are two
faces of the same granularity lever. The cost is a single offline sweep,
not per-shape tuning: one \((N_c, K_c)\) pair serves all twelve shapes,
the source is unchanged, and every output stays bit-identical.

\begin{figure}[t]
\centering
\centering
\pandocbounded{\includegraphics[keepaspectratio,alt={Throughput of the proposed pre-packed kernel at the QKV shape (128, 2\{,\}048, 2\{,\}048) as the column-panel width N\_c shrinks from 512 (four panels, P-cluster only) to 64 (32 panels, both AMX blocks); every point is a median over eleven isolated invocations on macOS Tahoe 26.5.1, bit-exact. Finer panels expose more work to Grand Central Dispatch, and the kernel crosses both the BNNSMatMul and BNNS Graph medians (Tables 3, 4) as it reaches the second AMX block. The coarse N\_c = 512 sits below both baselines; the deployed N\_c = 64 sits above both.}]{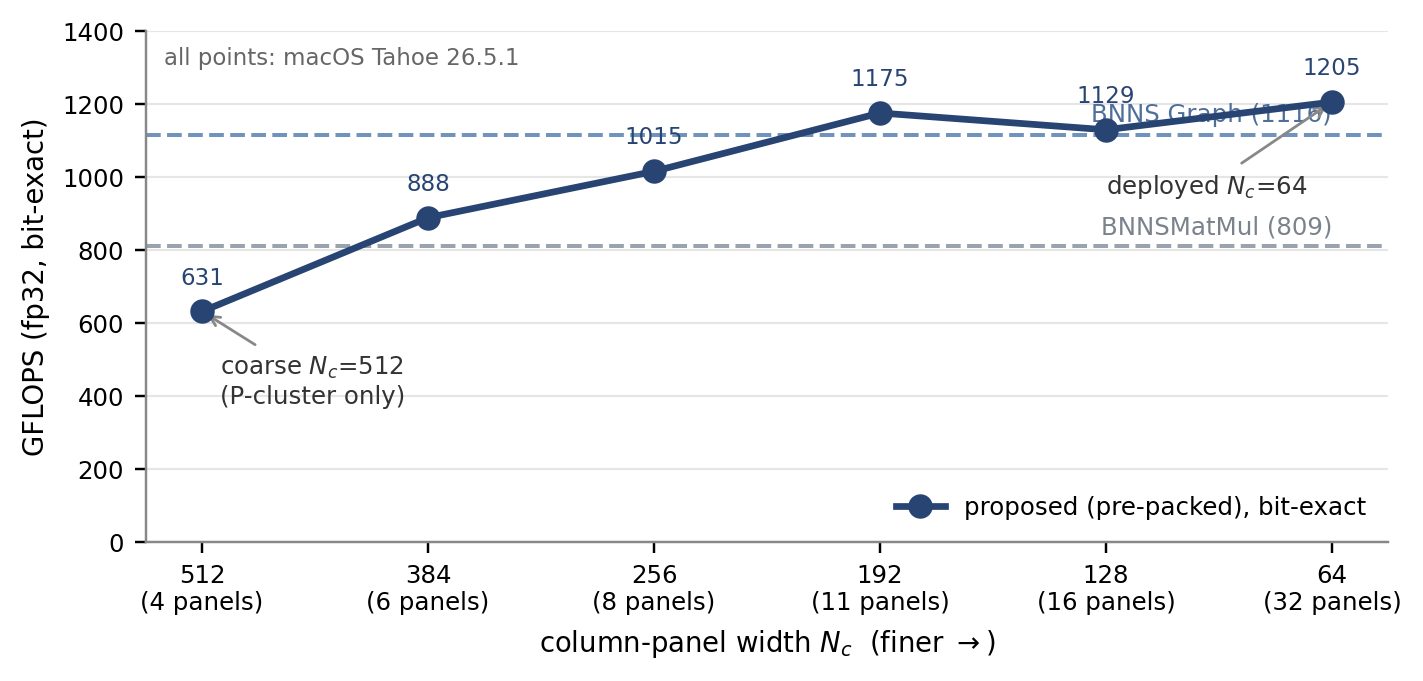}}
\caption{Throughput of the proposed pre-packed kernel at the QKV shape
\((128, 2{,}048, 2{,}048)\) as the column-panel width \(N_c\) shrinks
from 512 (four panels, P-cluster only) to 64 (32 panels, both AMX
blocks); every point is a median over eleven isolated invocations on
macOS Tahoe 26.5.1, bit-exact. Finer panels expose more work to Grand
Central Dispatch, and the kernel crosses both the BNNSMatMul and BNNS
Graph medians (Tables 3, 4) as it reaches the second AMX block. The
coarse \(N_c = 512\) sits below both baselines; the deployed
\(N_c = 64\) sits above both.}
\end{figure}

We measure the idle-block attribution directly rather than infer it from
the speedup. AMX has no public performance counter on the M1, but each
cluster's block can be driven only by an active core in that cluster, so
per-core CPU active residency is a sound proxy. Sampling per-core
residency (\passthrough{\lstinline!host\_processor\_info!}) across a
sustained loop of each backend, baseline-subtracted and reported as
active-core-equivalents out of four per cluster: at the QKV shape
cblas\_sgemm and BNNSMatMul hold the E-cluster at 0.00 -- they never
touch it -- and keep near a single P-core busy (0.4 to 1.3), the
near-single-thread occupancy matching their load-issue-bound floor; the
proposed kernel at \(N_c = 64\) drives roughly 3.0 to 3.5
P-core-equivalents and 1.9 to 2.5 on the E-cluster, engaging both
blocks. The same holds at the N-large FFN1 shape. The E-cluster
residency is 0.00 for both Accelerate routines at both shapes across six
runs, so the idle second block is an observed property of these shapes,
not only the most parsimonious account of the speedup. The cluster split
was identified by the standard trick confining
\passthrough{\lstinline!QOS\_CLASS\_BACKGROUND!} threads to the E-cores;
the probe is
\nolinkurl{bench/amx/amx\_core\_occupancy.cc}.

Two further parallelism choices were rejected (Table 1). Partitioning
the output over \(M\) as a second axis -- splitting the 128 rows into
16- or 32-row blocks -- lowers throughput (AMX wants the full
\(M = 128\) per work item), so the column panel is the only parallelism
axis that helps. And explicitly pinning the thread count, rather than
letting Grand Central Dispatch place work across the
\(\lceil N/N_c \rceil\) panels, stays within run-to-run noise at
\(N_c = 64\), so the default scheduler is left in place.

\subsubsection{4.7 End-to-end prefill GEMM
measurement}\label{end-to-end-prefill-gemm-measurement}

We measure end-to-end prefill GEMM time directly rather than composing
it from the per-shape medians. A standalone harness
(\nolinkurl{bench/amx/amx\_e2e\_measured.cc}) runs the
full prefill GEMM sequence of each model in layer order at \(S = 128\),
fp32: per transformer block the Q, K, and V projections, the
attention-output projection, the FFN up-projection, and the FFN
down-projection, over \(L\) blocks, then a single LM head. Attention
SDPA (\(Q \cdot K^{\top}\), softmax, the value sum) is not a GEMM in
this paper's sense and is omitted; the reported quantity is prefill GEMM
time, not full forward time. For the proposed backend every weight is
pre-packed once at load (untimed); each call then pays the activation
transpose and runs the compute loop against the resident panels. Unlike
the per-shape harness of Table 3, this activation transpose -- shared
only across the Q/K/V trio that consume the same input -- is inside the
timed region, so the measurement is strictly more conservative than a
FLOP-weighted composition. cblas\_sgemm and BNNSMatMul re-pack the
weight on every call as before. Each backend's full-sequence time is the
median over eleven isolated invocations (median of fifteen timed passes
within each).

\begin{table*}[t]
\centering
\caption*{Table 6. Measured end-to-end prefill GEMM time at \(S = 128\)
on the M1, median over eleven isolated invocations, weights pre-packed
once at load for the proposed kernel. \(H\) is the hidden dimension,
\(F\) the FFN intermediate dimension, \(V\) the vocabulary size, \(L\)
the number of blocks. The two rightmost columns are the proposed
kernel's speedup over each re-packing baseline. Source:
\protect\texttt{bench/amx/amx\_e2e\_measured.cc}.}
\footnotesize
\setlength{\tabcolsep}{4pt}
\begin{tabular}{@{}
>{\raggedright\arraybackslash}p{(\linewidth - 18\tabcolsep) * \real{0.1000}}
  >{\raggedright\arraybackslash}p{(\linewidth - 18\tabcolsep) * \real{0.1000}}
  >{\raggedright\arraybackslash}p{(\linewidth - 18\tabcolsep) * \real{0.1000}}
  >{\raggedright\arraybackslash}p{(\linewidth - 18\tabcolsep) * \real{0.1000}}
  >{\raggedright\arraybackslash}p{(\linewidth - 18\tabcolsep) * \real{0.1000}}
  >{\raggedright\arraybackslash}p{(\linewidth - 18\tabcolsep) * \real{0.1000}}
  >{\raggedright\arraybackslash}p{(\linewidth - 18\tabcolsep) * \real{0.1000}}
  >{\raggedright\arraybackslash}p{(\linewidth - 18\tabcolsep) * \real{0.1000}}
  >{\raggedright\arraybackslash}p{(\linewidth - 18\tabcolsep) * \real{0.1000}}
  >{\raggedright\arraybackslash}p{(\linewidth - 18\tabcolsep) * \real{0.1000}}@{}}

\toprule
\begin{minipage}[t]{\linewidth}\raggedright
model
\end{minipage} & \begin{minipage}[t]{\linewidth}\raggedright
H
\end{minipage} & \begin{minipage}[t]{\linewidth}\raggedright
F
\end{minipage} & \begin{minipage}[t]{\linewidth}\raggedright
V
\end{minipage} & \begin{minipage}[t]{\linewidth}\raggedright
L
\end{minipage} & \begin{minipage}[t]{\linewidth}\raggedright
cblas (ms)
\end{minipage} & \begin{minipage}[t]{\linewidth}\raggedright
BNNS (ms)
\end{minipage} & \begin{minipage}[t]{\linewidth}\raggedright
ours (ms)
\end{minipage} & \begin{minipage}[t]{\linewidth}\raggedright
vs BNNS
\end{minipage} & \begin{minipage}[t]{\linewidth}\raggedright
vs cblas
\end{minipage} \\
\midrule

TinyLlama-1.1B & 2,048 & 5,632 & 32,000 & 22 & 432 & 341 & 240 & 1.42 &
1.80 \\
Llama-7B & 4,096 & 11,008 & 32,000 & 32 & 3,207 & 1,799 & 1,200 & 1.50 &
2.67 \\
\bottomrule
\end{tabular}
\end{table*}

The measured prefill GEMM speedup over BNNSMatMul is \(1.42\times\) at
TinyLlama and \(1.50\times\) at Llama-7B, and over cblas\_sgemm -- the
path engines like llama.cpp call -- \(1.80\times\) and \(2.67\times\).
These wall-clock figures sit just below a FLOP-weighted composition of
the corresponding per-shape ratios from Table 3 (\(1.58\times\) for
TinyLlama and \(1.62\times\) for Llama-7B over BNNSMatMul, each weighted
by that model's own shapes -- not the all-twelve geometric mean) because
the activation transpose, hoisted out of the per-shape timing, is paid
inside every call here; it is the cost of the row-major pre-packed
layout and is included for honesty. The GPT-2-small architecture
(\(H = 768\)) is omitted: its shapes (\(N, K \le 3{,}072\)) lie below
the \(H \ge 2{,}048\) configurations measured here.

These are prefill GEMM times, not full forward times. To measure the
full-forward effect inside a deployed engine, we integrate the kernel
into llama.cpp {[}8{]}. Its BLAS backend
(\passthrough{\lstinline!ggml-blas!}) routes every prefill weight matmul
through a single \passthrough{\lstinline!cblas\_sgemm(NoTrans, Trans)!}
call whose \passthrough{\lstinline!src0!} is the constant fp32 weight
\(W[N, K]\); we intercept exactly that call for the constant-weight fp32
case, pre-packing each weight once at first use (Section 3.3) and
dispatching the multi-threaded \(N_c{=}64\), \(K_c{=}2{,}048\) kernel,
and fall back to \passthrough{\lstinline!cblas\_sgemm!} for everything
else (attention scores, small shapes). A single environment variable
toggles the routine, so the \emph{same binary} runs both arms over the
same fp32 GGUF -- isolating the GEMM routine as the only deliberate
difference. We use a TinyLlama-1.1B fp32 GGUF (4.1 GiB) so both arms see
identical weights; the quantized ggml path, where dequantization is
fused into the matmul, is a different code path our kernel does not
target. A first-call recomputation through
\passthrough{\lstinline!cblas\_sgemm!} confirms the substituted output
is bit-identical (maximum absolute difference \(0\)).

At a 128-token prefill (\passthrough{\lstinline!llama-bench!}, four
threads, the prefill length of every shape in this paper) the kernel
raises steady-state throughput from \(291\) to \(420\) tokens per
second, a measured \(1.44\times\) full-forward speedup (median over
twenty iterations with the cold first one excluded; coefficient of
variation 4.7 percent baseline, 5.3 percent ours). That first iteration
is excluded from both arms because it carries the one-time packing of
all 107 weight matrices the shim intercepts and the Grand Central
Dispatch thread-pool spin-up, costs paid once at model load in
deployment, not per token (shim and reproduction recipe:
\nolinkurl{bench/amx/llamacpp\_shim}). GEMM dominates this
fp32 prefill: the baseline full-forward time (roughly 440 ms per
128-token prefill at 291 tokens per second) is close to the 432 ms cblas
prefill GEMM time measured in the standalone harness of Table 6 -- the
two figures come from different codebases, so the agreement is only
approximate -- leaving only a small non-GEMM remainder (SDPA,
normalizations, rotary embedding, KV-cache writes), which at \(S = 128\)
is cheap relative to the projection GEMMs. The \(1.44\times\) therefore
sits below the \(1.80\times\) GEMM-only ratio of Table 6 not because
non-GEMM work dilutes it but because the in-engine kernel itself runs
slower than in isolation. The arithmetic makes the gap explicit: had the
in-engine kernel matched its isolated 240 ms (Table 6), the full-forward
time would fall to roughly 248 ms (\(\approx 516\) tokens per second, a
\(1.77\times\) gain) given the small non-GEMM remainder; the measured
305 ms (420 tokens per second) shows the in-engine kernel costing about
57 ms more than in isolation. That excess is integration overhead -- its
Grand Central Dispatch threads now contend with llama.cpp's own operator
thread pool, and the activation transpose is paid on every call rather
than shared across the QKV trio -- not a limit of the kernel, and the
isolated harness avoids both. The advantage is specific to the
prefill-length regime this paper characterizes: it narrows to
\(1.10\times\) at 256 tokens and reverses to \(0.82\times\) at 512,
where the per-call activation transpose (the weights are pre-packed, the
activations are not, and the transpose grows with \(M\)) and
Accelerate's stronger large-\(M\) scaling overtake the gain. A
serving-stack integration that hoists the activation transpose and
covers the quantized path is left to follow-up work.

\subsection{5. Related Work}\label{related-work}

\textbf{Direct programming of the Apple matrix path.} MpGEMM {[}1{]} is
the closest prior work: a hand-written cache-blocked SME kernel on the
M4 Pro, a 1.23 geometric mean over Accelerate across DeepSeek and LLaMA,
using BLIS-style partitioning, on-the-fly transposition, and specialized
microkernels. The present work complements it in three respects. First,
it targets AMX on the M1, which with the M2 and M3 (all lacking SME)
constitutes the majority of deployed Apple Silicon. Second, MpGEMM wins
inside the inner loop -- its SME multi-vector loads (up to 900 GB/s
against 230 for single-register loads) relieve the load bottleneck
within the microkernel; AMX lacks that and its inner loop is load-issue
bound (Section 4.2), so on the M1 the gain comes from above it: the
multi-thread scheduling that fills both blocks and weight pre-packing
(Section 4.4). Third, the present work bounds the design space with an
exhaustive negative-results set (Section 3.4) and reports a bit-exact
result, with its micro-architectural basis, on the more widely deployed
hardware.

\textbf{Weight pre-packing and BNNS Graph.} Packing a constant weight
once and reusing it is standard: XNNPACK caches repacked weights
{[}14{]}, oneDNN pre-packs into blocked layouts {[}15{]}, and Apple's
BNNS Graph (macOS 15) repacks at graph-compile time {[}16{]}. The
proposed kernel exceeds BNNS Graph at all twelve shapes (geometric mean
1.17) and stays bit-exact, while BNNS Graph reaches its throughput on
nine rectangular shapes at reduced precision; it is also a graph
compiler, not a callable GEMM, so using it means serializing and
compiling ahead of time.

\textbf{Multi-threaded AMX and the second block.} That the M1 carries
one AMX block per cluster is documented in the reverse-engineered
references {[}5, 6{]}, and we measure the two-block aggregate directly
(Table 5); Hübner et al.~{[}3{]} characterize Accelerate's multi-thread
throughput on the M-series more broadly. We do not claim that engaging
both blocks raises throughput as a novel mechanism; our observation is
that Accelerate's GEMM routines do not take it at LLM prefill shapes
(\(M = 128\)) -- a kernel whose weight panels are fine enough to fill
both blocks exceeds them even without pre-packing at the \(K \geq N\)
shapes (Section 4.4) -- and a direct per-core occupancy measurement
confirms the cause (Accelerate holds the E-cluster block at 0.00,
Section 4.6). The contribution is this measured, shape-resolved
attribution.

\textbf{Apple AMX references and custom kernels.} The corsix/amx {[}5{]}
and dougallj {[}6{]} repositories provide the canonical
reverse-engineered encodings, and philipturner/amx-benchmarks {[}9{]}
reports peak throughput; our vendored
\nolinkurl{third\_party/amx/aarch64.h} matches the corsix
HEAD and the encodings were verified against the corsix emulator.
Gazzoni Filho et al.~{[}17{]} hand-write direct-AMX kernels for
post-quantum cryptography (Saber, FrodoKEM) on the M1 and M3, evidence
that direct AMX programming is practical outside ML; their kernels are
single-threaded with batching and target a different arithmetic domain.
Zhou {[}2{]} develops an in-place masked-outer-product GEMM that exceeds
Accelerate on general shapes, using overlapping tiles to avoid
Accelerate's scratch buffers. We do not implement that technique; the
present work instead applies Goto--BLIS blocking {[}7{]} with explicit
pre-packing and per-shape \((K_c, N_c)\) tuning. Whether masked outer
products compose with the deployment levers studied here -- weight
pre-packing and multi-thread block-filling -- or improve on the present
kernel is an open question we leave to future work. Distinct from the
fp32 focus here, the sparse-ternary AMX GEMM of Lipshitz et al.~{[}10{]}
targets quantized rather than fp32 prefill -- a complementary
reduced-precision direction outside the scope of this comparison.

\textbf{LLM inference characterization on Apple Silicon.} Benazir and
Lin {[}11{]} characterize LLM inference on the M2 Ultra/Max and M4 Pro
at the 8B-405B scale across 14 quantization schemes, focused on the GPU
and unified-memory paths; a MLX-versus-PyTorch study {[}12{]} addresses
an adjacent question but not the CPU + AMX path with operator-level
attribution; the Bare-Metal study {[}4{]} builds a NEON engine treating
AMX as a black box (its NEON decode trails the Accelerate/AMX baseline
by 4.87\(\times\), derived from their reported throughputs). Catalán et
al.~{[}18{]} characterize Apple's CPU, GPU, AMX, and Neural Engine
across the M1 through M4 with GEMM, finding AMX the most efficient fp32
engine and noting the M4 Pro's two matrix accelerators, but measure the
vendor paths rather than building a kernel that exceeds them. Inspection
of the ONNX Runtime MLAS source {[}13{]} confirms it uses NEON kernels
rather than Accelerate on Arm64, so an ONNX Runtime single-thread
prefill comparison is against an OpenBLAS-class NEON baseline.

\subsection{6. Conclusion}\label{conclusion}

This paper examined where, and why, a hand-written prefill GEMM kernel
can exceed Apple's Accelerate on the M1 AMX coprocessor, and reached a
structural rather than a tuning conclusion. The M1 AMX inner loop is
load-issue bound: with any operand load interleaved with the FMA32
stream a single thread drops into roughly the 610-to-680 GFLOPS band,
well under half the load-free rate, regardless of tile shape, load
pairing, load count, the load-to-FMA dependency, or phase batching, and
no microkernel rearrangement we evaluate escapes it. Across several
threads the two on-chip AMX blocks deliver an aggregate near 1,480
GFLOPS, which the deployed kernel reaches only when its column panels
are fine enough to keep every core fed.

The advantage over Accelerate therefore comes from two levers above the
inner loop, not a faster kernel: fine multi-thread panels fill the
second AMX block Accelerate leaves idle (winning the \(K \geq N\)
shapes), and pre-packing the constant weight wins the \(N > K\) shapes.
Both levers are known in isolation; what the kernel-centric framing of
prior work does not provide -- and what this paper contributes -- is the
measurement that Accelerate forgoes both at LLM prefill shapes, resolved
by shape class and confirmed by a direct per-core occupancy probe
showing Accelerate hold the E-cluster block at zero while the proposed
kernel drives it (Section 4.6). A bit-exact kernel built on both exceeds
every Accelerate fp32 path we measure (cblas\_sgemm, BNNSMatMul, and
BNNS Graph at all twelve; BNNS-FC at the large-\(N\) shapes): a
geometric mean of 1.58 over BNNSMatMul, a larger margin (about
\(2.0\times\)) over cblas\_sgemm, and 1.17 over BNNS Graph -- bit-exact
at all twelve, where BNNS Graph is bit-exact at only three and computes
the rest at reduced precision. The negative results leave fine-panel
scheduling and weight pre-packing as the two levers that move
throughput, the inner loop otherwise at the hardware limit. This is the
bit-exact M1-AMX counterpart, on the more widely deployed hardware, to
MpGEMM's SME result.

\subsubsection{Limitations}\label{limitations}

The comparison fixes fp32: against cblas\_sgemm, BNNSMatMul, and BNNS
Graph the kernel is the fastest bit-exact fp32 path on the M1, but BNNS
Graph runs faster when allowed its default reduced precision, and that
headroom is unaddressed -- the M1 AMX fp16 path has roughly twice the
fp32 limit and is under-exploited by Accelerate's BNNS fp16 routines
(the int8 path is not a lever: the M1 vecint primitive runs at about
five cycles per instruction). The fp32 comparison we report is the
conservative one: at the three shapes where BNNS Graph also holds fp32
the kernel still leads (\(1.09\times\)), and the larger margins over
BNNS Graph arise only where it relaxes precision -- the advantage is
never bought with accuracy. The advantage requires the pre-packed-weight
deployment: for a single isolated GEMM the kernel is no faster than
BNNSMatMul (Section 4.4); the speedup is realized only when the packed
weight is reused across calls, as across the tokens and decode steps of
a serving session. The llama.cpp full-forward speedup of Section 4.7 is
measured wall-clock end-to-end (\(1.44\times\) at 128-token prefill, the
fp32 BLAS path), but it is confined to that path: a serving-stack
integration that also accelerates the quantized ggml path (fused
dequantization), hoists the per-call activation transpose, and extends
past the short-prefill regime where the gain holds is the primary
deferred validation. The result is for prefill (\(S = 128\),
compute-bound), not decode, whose single-token GEMV shapes (\(M = 1\))
are memory-bandwidth-bound. Finally, the two parts of the result
generalize differently. The load-issue bound on the inner loop (Section
4.2) is a property of the AMX FMA32/load issue port and is expected to
hold across the M1 through M3, which share the AMX microarchitecture.
The two-block aggregate (roughly 1,480 GFLOPS) and the panel width that
reaches it (Section 4.6) instead depend on the
four-P-core-plus-four-E-core, single-P-cluster organization measured
here -- which the base M2 and M3 share, but which the multi-cluster M2
Pro, Max, and Ultra do not, and which the SME-based M4 and later replace
entirely; on those parts the aggregate ceiling and the panel granularity
that reaches it would differ. All measurements are from a single M1 unit
(T8103) and we do not characterize these other parts directly.

\subsubsection{Future work}\label{future-work}

The most direct extension is an fp16 kernel on the same M1 AMX hardware:
whether an fp16-input, fp32-accumulate kernel escapes the load-issue
bound -- the FMA16 outer product performs four times the
multiply-accumulates per operand load -- together with an accuracy map
of where reduced-precision input is safe for prefill. A second is an SME
counterpart on the M4 and later, with MpGEMM as the baseline, to test
whether the structural account transfers to a unit whose multi-vector
loads relieve the very bottleneck identified here. A third is
integration into a deployed serving stack (for example the cblas\_sgemm
path of llama.cpp {[}8{]}) with end-to-end tokens-per-second reporting.

The implementation, evaluation benchmarks, the prior-art map at
\nolinkurl{docs/LITERATURE.md}, the investigation log at
\nolinkurl{docs/AMX\_REPORT.md}, and the reproduction
commands are available at \url{https://github.com/dbhan08/inferc}.

\subsection{References}\label{references}

{[}1{]} C. Deng, W. Yang, J. Fang, and D. Dong, ``Demystifying ARM SME
to Optimize General Matrix Multiplications,'' arXiv:2512.21473, December
2025.

{[}2{]} J. Zhou, ``Performance Analysis of the Apple AMX Matrix
Accelerator,'' M.Eng. thesis, Department of Electrical Engineering and
Computer Science, Massachusetts Institute of Technology, September 2025.
Available at
\url{https://commit.csail.mit.edu/papers/2025/Jonathan_Zhou_SB_Thesis.pdf}.

{[}3{]} P. Hübner, A. Hu, I. Peng, and S. Markidis, ``Apple vs.~Oranges:
Evaluating the Apple Silicon M-Series SoCs for HPC Performance and
Efficiency,'' arXiv:2502.05317, 2025.

{[}4{]} B. Kilictas and F. Alpay, ``Bare-Metal Tensor Virtualization:
Overcoming the Memory Wall in Edge-AI Inference on ARM64,''
arXiv:2601.03324, January 2026.

{[}5{]} P. Cawley (corsix), ``corsix/amx: reverse-engineered Apple AMX
instruction reference,'' GitHub repository,
\url{https://github.com/corsix/amx}, accessed May 2026.

{[}6{]} D. Johnson, ``dougallj: Apple silicon reverse-engineering
notes,'' GitHub repository, \url{https://github.com/dougallj}, accessed
May 2026.

{[}7{]} K. Goto and R. A. van de Geijn, ``Anatomy of high-performance
matrix multiplication,'' ACM Transactions on Mathematical Software,
vol.~34, no. 3, article 12, May 2008.

{[}8{]} G. Gerganov and the llama.cpp contributors, ``llama.cpp:
efficient LLM inference in pure C/C++,'' GitHub repository,
\url{https://github.com/ggml-org/llama.cpp}, accessed May 2026.

{[}9{]} P. Turner, ``amx-benchmarks: throughput measurements for the
Apple matrix coprocessor,'' GitHub repository,
\url{https://github.com/philipturner/amx-benchmarks}, accessed May 2026.

{[}10{]} B. Lipshitz, A. Melone, C. Maraziaris, and M. Bilal,
``Accelerating Sparse Ternary GEMM for Quantized ML on Apple Silicon,''
arXiv:2510.06957, 2025.

{[}11{]} A. Benazir and F. X. Lin, ``Benchmarking and Characterization
of Large Language Model Inference on Apple Silicon,'' Proceedings of the
ACM on Measurement and Analysis of Computing Systems (POMACS), vol.~9,
no. 3, article 48, pp.~1--26, 2 December 2025. DOI: 10.1145/3771563.

{[}12{]} O. A. Ajayi and O. Odunayo, ``Benchmarking On-Device Machine
Learning on Apple Silicon with MLX,'' presented at the 2024 Deep
Learning Indaba, Dakar, Senegal; arXiv:2510.18921, October 2025.

{[}13{]} Microsoft, ``ONNX Runtime: cross-platform inference engine,''
GitHub repository, \url{https://github.com/microsoft/onnxruntime},
accessed May 2026. MLAS Arm64 SGEMM kernels at
\nolinkurl{onnxruntime/core/mlas/lib/aarch64/SgemmKernelNeon.S}
and the NEON/DOT/I8MM/SVE dispatch in
\nolinkurl{onnxruntime/core/mlas/lib/platform.cpp}.

{[}14{]} Google, ``XNNPACK: high-efficiency neural-network inference
operators,'' GitHub repository, \url{https://github.com/google/XNNPACK};
and ``Memory-efficient inference with XNNPACK weights cache,''
TensorFlow Blog, June 2022. (XNNPACK repacks and caches constant weights
once for reuse.)

{[}15{]} Intel, ``oneDNN: weight pre-packing for matmul and
inner-product primitives,'' oneAPI Deep Neural Network Library
documentation, accessed May 2026. (Constant weights are converted once
to a blocked layout and reused, avoiding per-call repacking.)

{[}16{]} Apple, ``What's new in BNNS Graph,'' WWDC25 session 276, and
``Support real-time ML inference on the CPU,'' WWDC24 session 10211,
\url{https://developer.apple.com/videos/play/wwdc2025/276/}. (BNNS
Graph, macOS 15+, automatically repacks weights for cache locality.)

{[}17{]} D. L. Gazzoni Filho, G. Brandão, G. Adj, A. Alblooshi, I. A.
Canales-Martínez, J. Chávez-Saab, and J. López, ``PQC-AMX: Accelerating
Saber and FrodoKEM on the Apple M1 and M3 SoCs,'' in 2024 IEEE 31st
Symposium on Computer Arithmetic (ARITH), pp.~9--16, 2024; IACR ePrint
2024/195. (Single-threaded hand-written direct-AMX kernels for the Saber
and FrodoKEM post-quantum schemes on the Apple M1 and M3.)

{[}18{]} S. Catalán, R. Rodríguez-Sánchez, C. García Sánchez, and L.
Piñuel Moreno, ``A comparative performance and efficiency analysis of
Apple's M architectures: A GEMM case study,'' Future Generation Computer
Systems, vol.~180, article 108393, 2026. DOI:
10.1016/j.future.2026.108393.

\end{document}